\documentclass[graybox]{svmult}
\usepackage{graphics,lscape,graphicx,epsfig,amsmath,amssymb}
\usepackage{mathptmx}       
\usepackage{helvet}         
\usepackage{courier}        
\usepackage{type1cm}        
\usepackage{makeidx}         
\usepackage{graphicx}        
\usepackage{multicol}        
\usepackage[bottom]{footmisc}

\makeindex             

\allowdisplaybreaks

\newcommand{\be}{\begin{equation}}
\newcommand{\ee}{\end{equation}}
\newcommand{\bea}{\begin{eqnarray}}
\newcommand{\eea}{\end{eqnarray}}
\newcommand{\ba}{\begin{eqnarray*}}
\newcommand{\ea}{\end{eqnarray*}}
\newcommand{\dagga}{{\phantom{\dagger}}}

\newcommand{\dis}{\displaystyle}

\newcommand{\up}{\uparrow}
\newcommand{\down}{\downarrow}
\newcommand{\fract}[2]{\frac{\dis #1}{\dis #2}}
\newcommand{\Tr}{\mathrm{Tr}}
\newcommand{\eqn}[1]{(\ref{#1})}
\begin{document}

\title*{The Out-of-Equilibrium Time-Dependent Gutzwiller Approximation}
\author{Michele Fabrizio}
\institute{Michele Fabrizio \at International School for Advanced Studies, SISSA, via Bonomea 265, I-34136, Trieste, Italy,
and The Abdus Salam Center for Theoretical Physics, ICTP,  P.O. Box 586, 34100, Trieste, Italy. \email{fabrizio@sissa.it}}
\maketitle
\abstract{We review the recently proposed extension of the Gutzwiller approximation, M. Schir\`o and M. Fabrizio, Phys. Rev. Lett. {\bf 105}, 076401 (2010), designed to describe the out-of-equilibrium time-evolution of a Gutzwiller-type variational wave function for correlated electrons. The method, which is strictly variational in the limit of infinite lattice-coordination, is quite general and flexible, and it is applicable to generic non-equilibrium conditions, even far beyond the linear response regime. 
As an application, we discuss the quench dynamics of a single-band Hubbard model at half-filling, 
where the method predicts a dynamical phase transition above a critical quench that resembles the sharp 
crossover observed by time-dependent dynamical mean field theory. 
We next show that one can actually define in some cases a multi-configurational wave function combination 
of a whole set of mutually orthogonal Gutzwiller wave functions. The Hamiltonian projected in that subspace can be exactly evaluated and is equivalent to a model of auxiliary spins coupled to 
non-interacting electrons, closely related to the slave-spin theories for correlated electron models. 
The Gutzwiller approximation turns out to be nothing but the mean-field approximation applied to that 
spin-fermion model, which displays, for any number of bands and integer fillings,  
a spontaneous $Z_2$ symmetry breaking that can be identified as the Mott insulator-to-metal transition. 
}
\section{Introduction}
Time-resolved spectroscopies are advancing incredibly fast towards accessing ultra-short time ($\lesssim$ femtoseconds) dynamics.\cite{Giannetti_NatComm11,Ichikawa_NatMat11,Fausti_Science11,
RevModPhys.74.895,Attosecond_RMP} On such timescales, it becomes possible to monitor 
how the electronic degrees of freedom 
react to a sudden external stimulus before electrons have time to equilibrate with the lattice, which commonly starts after few picoseconds. In this initial transient regime, one can therefore neglect the coupling to the lattice and study just the way how collisions among the electrons brought about by interaction redistribute the excess energy injected into the system.  This situation in which the electrons provide their own dissipative bath has recently attracted interest especially in connection with cold atoms trapped in optical lattices,\cite{Silva-RMP} which realize systems where the particles are, to a large extent, 
ideally isolated from the environment. 
There are by now several claims that, when correlation is strong enough and the injected energy exceeds a threshold, the electrons alone are unable to efficiently exchange energy by collisions, hence remain trapped for long time into non-thermal configurations. The most convincing evidences come from dynamical mean field theory  (DMFT) simulations of quantum quenches in the half-filled single-band Hubbard model.\cite{Werner_prl09,Werner-PRB} Such a technique is however computationally heavy and does not allow accessing very long times. 
Alternatively, qualitatively similar results have been reproduced by a much simpler tool, the time-dependent Gutzwiller approximation (t-GA),\cite{Marco-PRL,Marco-PRB} which allows to follow much longer the time evolution, although it lacks enough dissipative channels to describe the system flowing towards a steady state.\cite{Marco-PRB}  Nevertheless, the time averages of the observables as obtained 
through t-GA agree satisfactorily with the DMFT steady state values, which justifies using t-GA as a valid  alternative to more sophisticated approaches, like DMFT, for its simplicity and flexibility. 

Here, we shall present in detail how t-GA can be implemented efficiently in a generic multi-band lattice model of electrons mutually coupled by a short-range interaction. We will show that the method is able to access the full out-of-equilibrium dynamics also far beyond the linear response regime discussed in 
Ref.~\cite{Lorenzana}. In particular, a nice feature of t-GA is its ability of treating on equal footing 
the dynamics both of the low-energy coherent quasiparticles as well as of the high-energy incoherent excitations, which are commonly refereed to as the Hubbard side-bands close to the Mott transition. 
Within t-GA these two distinct excitations, quasiparticles and Hubbard bands, possess their own dynamics, and influence each other only in a mean-field like fashion. This is clearly an approximation of the 
actual time evolution, and the reason why the method lacks enough dissipation, although the 
ensuing dynamics is much richer than the conventional time-dependent Hartree-Fock. 

Finally, we discuss some instructive connections between t-GA 
and the recently developed slave-spin representations of the 
Hubbard model.\cite{DeMedici-1,DeMedici-2,Z2-1,Z2-2} Essentially, we will show that in the limit 
of infinite lattice-coordination, where the Gutzwiller approximation becomes an exact variational approach, and under particular circumstances, e.g. integer filling in a multi-band model, one can actually define a multi-configurational basis of Gutzwiller wave-functions and explicitly evaluate the Hamiltonian matrix elements. It turns out that the Hamiltonian projected onto that basis coincides with its slave-spin representation with the major advantage that the constraint required in the slave-spin theory to project the enlarged Hilbert space onto the physical one can be here enforced exactly.  

\section{The model and the Gutzwiller wavefunction and approximation}
\label{Model}

We shall consider the following tight-binding model on a lattice with 
coordination number $z$:  
\be
\mathcal{H} = \sum_{i,j}\sum_{a,b=1}^N\,\Big(
t^{ab}_{ij}\, c^\dagger_{ia}c^\dagga_{jb}+\text{H.c.}\Big) + 
\sum_i \, \mathcal{U}_i,
\label{Ham}
\ee
where $c^\dagger_{ia}$ creates an electron at site $i$ in orbital $a=1,\dots,N$, the index $a$ including also the spin,  and $\mathcal{U}_i$ is a local term 
that accounts also for the interaction. The hopping parameter $t^{ab}_{ij}$ is assumed to scale like 
$1/z^{r/2}$ where $r$ is the lattice distance between sites $i$ and $j$, so that the average hopping energy per site remains finite also in the limit $z\to\infty$.\cite{DMFT} 
The Gutzwiller wavefunction\cite{Gutzwiller_1,Gutzwiller_2} is defined through
\be
\mid\Psi\rangle = \mathcal{P}\mid\Psi_0\rangle = \prod_i\,\mathcal{P}_i\mid\Psi_0\rangle,
\label{Psi}
\ee
where $\mid\Psi_0\rangle$ is a Slater determinant\footnote{In reality, for the method to work it is enough 
that Wick's theorem applies, hence $\mid \Psi_0\rangle$ could even be a BCS wavefunction. Here, for 
sake of simplicity, we shall only consider Slater determinants.}
and $\mathcal{P}_i$ a local operator that we will denote, although improperly, as the {\sl Gutzwiller projector}, whose role is 
to the change the weights of the local electronic configurations with respect to the Slater determinant. Both $\mid \Psi_0\rangle$ and $\mathcal{P}_i$ have to be determined variationally to minimize the total energy
\be
E = \fract{\langle \Psi\mid \mathcal{H}\mid \Psi\rangle}
{\langle \Psi\mid\Psi\rangle}.\label{total-E}
\ee

The Guzwiller approximation begins by imposing, for reasons that will become clear soon, the following two constraints on $\mathcal{P}_i$:\cite{mio-dimero}
\bea
\langle \Psi_0\mid \mathcal{P}_{i}^\dagger \mathcal{P}_{i}^\dagga\mid \Psi_0\rangle &=& 1,\label{1}\\
\langle \Psi_0\mid \mathcal{P}_{i}^\dagger \mathcal{P}_{i}^\dagga
\,c^\dagger_{ia}c^\dagga_{ib}\mid \Psi_0\rangle &=& \langle \Psi_0\mid 
c^\dagger_{ia} c^\dagga_{ib}\mid \Psi_0\rangle, \qquad \forall a,b.\label{2}
\eea
These constraints mean that, if we select from the operator $\mathcal{P}_i^\dagger \mathcal{P}_i^\dagga$ any two fermionic operators and average over the Slater determinant what remains, then such an average vanishes identically. This property is very convenient if the lattice coordination $z$ tends to infinity. In fact, we note that, for $i\not = j$,  
\bea
\langle \Psi_0 \mid   \mathcal{P}_i^\dagger \mathcal{P}_i^\dagga\,
\mathcal{P}_j^\dagger \mathcal{P}_j^\dagga\mid \Psi_0\rangle &=& 
\langle \Psi_0 \mid   \mathcal{P}_i^\dagger \mathcal{P}_i^\dagga\mid\Psi_0\rangle
\langle\Psi_0\mid \mathcal{P}_j^\dagger \mathcal{P}_j^\dagga\mid \Psi_0\rangle \nonumber\\
&& + \langle \Psi_0 \mid   \mathcal{P}_i^\dagger \mathcal{P}_i^\dagga\,
\mathcal{P}_j^\dagger \mathcal{P}_j^\dagga\mid \Psi_0\rangle_{\text{connected}}\nonumber\\
&& =  1 + \langle \Psi_0 \mid   \mathcal{P}_i^\dagger \mathcal{P}_i^\dagga\,
\mathcal{P}_j^\dagger \mathcal{P}_j^\dagga\mid \Psi_0\rangle_{\text{connected}},
\label{connected}
\eea
where the last term on the right hand side includes all Wick's contractions connecting the two sites, 
and the constant 1 comes from \eqn{1}. Because of the constraint \eqn{2}, the terms that connect 
the two sites by only two fermionic lines vanish, leaving only terms with $2n>2$  connecting lines. 
In the limit of infinite lattice-coordination, these latter terms vanish like $z^{-nR_{ij}}$, where 
$R_{ij}$ is the minimum length of the path connecting $i$ to $j$. For a given $i$, if we consider all 
sites $j$ at fixed $R_{ij}=R$ and sum over them Eq.~\eqn{connected}, each connected term above will contribute $\sim z^{-nR}$, $n>1$, but there are only $\sim z^R$ such terms so that, 
in the limit $z\to\infty$, their sum will vanish. 
This property simplifies considerably all calculations in the infinite lattice-coordination limit, which we 
shall assume hereafter. In particular, it implies that\cite{mio-dimero,Gebhard} 
\[
\langle \Psi \mid\Psi\rangle = \prod_i \langle \Psi_0\mid \mathcal{P}_i^\dagger \mathcal{P}_i^\dagga
\mid \Psi_0\rangle =1,
\]
namely the wavefunction \eqn{Psi} is normalized, and moreover that, given any local operator $\mathcal{O}_i$, 
\be
\langle \Psi\mid \mathcal{O}_i\mid\Psi\rangle = 
\langle \Psi_0\mid \mathcal{P}_{i}^\dagger \mathcal{O}_i \mathcal{P}_i^\dagga\mid\Psi_0\rangle,
\label{O_i}
\ee
which can be easily evaluated by Wick's theorem. In addition, it also follows that 
\be
\sum_{i,j}\, t^{ab}_{ij}\,\langle \Psi\mid c^\dagger_{ia}c^\dagga_{jb}\mid\Psi\rangle = 
\sum_{i,j}\, t^{ab}_{ij}\,\langle \Psi_0\mid \mathcal{P}_{i}^\dagger c^\dagger_{ia} \mathcal{P}_i^\dagga
\, \mathcal{P}_{j}^\dagger c^\dagga_{jb} \mathcal{P}_j^\dagga
\mid\Psi_0\rangle,
\label{C_ij}
\ee
where one has to keep only Wick's contractions that connect sites $i$ and $j$ by just a single fermionic line, 
since the terms with three or more lines vanish in the limit 
$z\to\infty$. A simple way to proceed is by defining the matrix elements $R_{i\,ab}$ through 
\be
\langle \Psi_0 \mid 
\mathcal{P}_i^\dagger c^\dagger_{ia}
\mathcal{P}_i^\dagga c^\dagga_{ic} \mid \Psi_0\rangle 
\equiv  \sum_c\, R^*_{i\,ab} \,\langle \Psi_0\mid c^\dagger_{ib}c^\dagga_{ic}\mid\Psi_0\rangle, 
\label{def:R_ab}
\ee
that automatically include all Wick's contractions after extracting from 
the operator $\mathcal{P}_i^\dagger c^\dagger_{ia}
\mathcal{P}_i^\dagga$ a single fermionic line. Through \eqn{def:R_ab} 
we can formally write Eq.~\eqn{C_ij} as 
\be
 \langle \Psi\mid c^\dagger_{ia}c^\dagga_{jb}\mid\Psi\rangle = 
 \sum_{cd}\, R^\dagger_{i\,ca} R^\dagga_{j\, bd}
\,\langle \Psi_0\mid c^\dagger_{ic}c^\dagga_{jd}\mid\Psi_0\rangle.\label{def:hopping}
\ee

In conclusion, provided \eqn{1} and \eqn{2} are satisfied, and upon 
defining through Eq.~\eqn{def:R_ab} the renormalized hopping amplitude
\be
t^{ab}_{*\,ij} \equiv  \sum_{cd}\, R^\dagger_{i\,ac}\,t^{cd}_{ij}\,
R^\dagga_{j\,db},\label{def:t}
\ee
and the non-interacting Hamiltonian 
\be
\mathcal{H}_* =  \sum_{i,j}\,\sum_{ab}\,\Big( 
t^{ab}_{*\,ij}\,c^\dagger_{ia} 
c^\dagga_{jb} + H.c.\Big),\label{def:H*}
\ee
then the average energy in the limit of infinite lattice coordination is 
\be
E =  \langle\Psi_0\mid\mathcal{H}_*\mid \Psi_0\rangle 
+ \sum_i\,\langle\Psi_0\mid \mathcal{P}^\dagger_i
\mathcal{U}_i \mathcal{P}^\dagga_i \mid \Psi_0\rangle,\label{total-E-final}
\ee
which can be evaluated by Wick's theorem. Minimization of 
\eqn{total-E-final} with respect to all variational parameters 
provides an estimate of the ground state energy. 
The expression \eqn{total-E-final}, with the definition \eqn{def:R_ab}, 
is strictly valid only in the limit of infinite lattice-coordination. However, it is common to keep using the same expressions also for finite-coordination lattices, hence the name {\sl Gutzwiller approximation}.

Like any other variational approach, also the one we just outlined can only provide information on static properties, assumed to represent well those of the actual ground state. Here we shall propose an extension that allows to access also dynamical properties.\cite{Marco-PRL,Marco-PRB}

\section{Time-dependent Gutzwiller approximation}
\label{Time-dependent Gutzwiller approximation}

From now on we shall assume that both the Slater determinant as well as the Gutzwiller projectors are time-dependent, hence
\be
\mid \Psi(t)\rangle = \mathcal{P}(t)\mid \Psi_0(t)\rangle =
\prod_i \mathcal{P}_i(t)\mid \Psi_0(t)\rangle.\label{tGA-Psi}
\ee
If the Eqs.~\eqn{1} and \eqn{2} are satisfied at any $t$, then, at 
any instant of time and in the limit of infinite coordination number, 
the average value of the Hamiltonian $E(t)$ will have the same expression 
as in Eq.~\eqn{total-E-final}, i.e.
\be
E(t) =  \langle\Psi_0(t)\mid\mathcal{H}_*(t)\mid \Psi_0(t)\rangle 
+ \sum_i\,\langle\Psi_0(t)\mid \mathcal{P}_i(t)^\dagger\,
\mathcal{U}_i \,\mathcal{P}_i(t) \mid \Psi_0(t)\rangle.
\label{E(t)}
\ee
In particular, $\mathcal{H}_*(t)$ becomes time dependent since $R_{i\,ab}(t)$ depends on time. We shall adopt the variational principle that $\mid\Psi(t)\rangle$ is as close as possible to the solution of the 
Schr{\oe}dinger equation. Specifically,\cite{Marco-PRL} we define the functional 
$\mathcal{S}(t) = \int_0^t d\tau \,\mathcal{L}(\tau)$, that plays the role of a classical action, with Lagrangian
\bea
\mathcal{L}(t) &=& i\langle\Psi(t)\mid \dot{\Psi}(t)\rangle -E(t)
= i\langle\Psi_0(t)\mid \mathcal{P}(t)^\dagger 
\mathcal{P}(t)\mid \dot{\Psi}_0(t)\rangle \nonumber\\
&& +i \langle\Psi_0(t)\mid \mathcal{P}(t)^\dagger 
\dot{\mathcal{P}}(t)\mid\Psi_0(t)\rangle -E(t),
\label{tGA-L1}
\eea
and determine $\mid\Psi_0(t)\rangle$ and $\mathcal{P}_i(t)$  by the saddle point of the action 
under the two constraints Eqs.~\eqn{1} and \eqn{2}.   

Since $\mid\Psi_0(t)\rangle$ is a Slater determinant at any instant of time, then
\[
i\mid\dot{\Psi}_0(t)\rangle = \mathcal{V}(t)\mid\Psi_0(t)\rangle,
\]
with 
\[
\mathcal{V}(t)=\sum_i\,\mathcal{V}_i(t) 
+ \sum_{i\not =j}\,\mathcal{V}_{ij}(t),
\]
a single-particle operator that contains local terms $\mathcal{V}_i(t)$ as well as hopping terms $\mathcal{V}_{ij}(t)$. We note that, 
because of Eqs.~\eqn{1} and \eqn{2}, it follows that 
\bea
\langle \Psi_0(t)\mid \mathcal{P}(t)^\dagger 
\mathcal{P}(t)^\dagga \,\mathcal{V}_i(t) \mid \Psi_0(t)\rangle 
&=& \langle \Psi_0(t)\mid \mathcal{P}_i(t)^\dagger 
\mathcal{P}_i(t)^\dagga \,\mathcal{V}_i(t) \mid \Psi_0(t)\rangle \nonumber\\
&=& \langle \Psi_0(t)\mid \mathcal{V}_i(t) \mid \Psi_0(t)\rangle. 
\label{V1}
\eea
Seemingly, 
\ba
&&\langle \Psi_0(t)\mid \mathcal{P}(t)^\dagger 
\mathcal{P}(t)^\dagga \,\mathcal{V}_{ij}(t) \mid \Psi_0(t)\rangle \\
&& = \langle \Psi_0(t)\mid \mathcal{P}_i(t)^\dagger 
\mathcal{P}_i(t)^\dagga \,\mathcal{P}_j(t)^\dagger 
\mathcal{P}_j(t)^\dagga \,\mathcal{V}_{ij}(t) \mid \Psi_0(t)\rangle 
= \langle \Psi_0(t)\mid \mathcal{V}_{ij}(t) \mid \Psi_0(t)\rangle\\
&& ~~~~+\langle \Psi_0(t)\mid \mathcal{P}_i(t)^\dagger 
\mathcal{P}_i(t)^\dagga \,\mathcal{P}_j(t)^\dagger 
\mathcal{P}_j(t)^\dagga \,\mathcal{V}_{ij}(t) 
\mid \Psi_0(t)\rangle_{\text{connected}}.
\ea
The {\sl connected} term on the right hand side means that we have 
to extract out of $\mathcal{P}_i(t)^\dagger 
\mathcal{P}_i(t)^\dagga$ a number of fermionic operators, which are to be multiple of two, one of which has to be contracted with $\mathcal{V}_{ij}(t)$, and the remaining ones with $\mathcal{P}_j(t)^\dagger 
\mathcal{P}_j(t)^\dagga$. By construction, the terms where we extract 
only two operators and average over $\mid \Psi_0(t)\rangle$ what remains, will vanish 
because of Eq.~\eqn{2}, while all the 
others, with four or more operators that are extracted, vanish in the limit of infinite coordination number. In conclusion, only the disconnect term survives, hence
\be
\langle \Psi_0(t)\mid \mathcal{P}(t)^\dagger 
\mathcal{P}(t)^\dagga \,\mathcal{V}_{ij}(t) \mid \Psi_0(t)\rangle 
= \langle \Psi_0(t)\mid \mathcal{V}_{ij}(t) \mid \Psi_0(t)\rangle,
\label{V2}
\ee
which, together with Eqs.~\eqn{V1}, imply that 
\bea
i\langle\Psi_0(t)\mid \mathcal{P}(t)^\dagger 
\mathcal{P}(t)\mid \dot{\Psi}_0(t)\rangle &=& 
\langle\Psi_0(t)\mid \mathcal{P}(t)^\dagger 
\mathcal{P}(t)^\dagga\,\mathcal{V}(t)\mid \Psi_0(t)\rangle
\nonumber\\
&=& \langle\Psi_0(t)\mid \mathcal{V}(t)\mid \Psi_0(t)\rangle
= i\langle\Psi_0(t)\mid \dot{\Psi}_0(t)\rangle.\label{V3}
\eea
Finally, Eqs.~\eqn{1} and \eqn{2} also lead to 
\be
i \langle\Psi_0(t)\mid \mathcal{P}(t)^\dagger 
\dot{\mathcal{P}}(t)^\dagga\mid\Psi_0(t)\rangle
= \sum_i\, i \langle\Psi_0(t)\mid \mathcal{P}_i(t)^\dagger 
\dot{\mathcal{P}}_i(t)\mid\Psi_0(t)\rangle.\label{V4}
\ee
As a result, Eq.~\eqn{tGA-L1} can be written as  
\bea
\mathcal{L}(t) 
&=& i\langle\Psi_0(t)\mid \dot{\Psi}_0(t)\rangle 
+i \sum_i\, \langle\Psi_0(t)\mid \mathcal{P}_i(t)^\dagger 
\dot{\mathcal{P}}_i(t)\mid\Psi_0(t)\rangle -E(t).
\label{tGA-L}
\eea

\subsection{A more convenient representation}
In order to make it easier the search for the saddle point, 
it is convenient to follow the method outlined in Ref.~\cite{Nicola-HF}, closely connected to the rotationally invariant slave-boson formalism of Ref.~\cite{Rotationally}. We assume there exists a set of creation and annihilation operators,  the natural basis operators $d^\dagger_{i\alpha}$ and $d^\dagga_{i\alpha}$, respectively, related to the original operators, 
$c^\dagger_{ia}$ and $c^\dagga_{ia}$, by a unitary transformation and such that 
\be
\langle \Psi_0(t)\mid d^\dagger_{i\alpha}d^\dagga_{i\beta}\mid\Psi_0\rangle = 
\delta_{\alpha\beta}\,n^0_{i\alpha}(t).\label{tGA-natural}
\ee
We introduce the Fock states in the natural basis 
\be
\mid i;\{n\}\rangle 
= \prod_\alpha \Big(d^\dagger_{i\alpha}\Big)^{n_{\alpha}}\mid 0\rangle, \label{tGA-Fock}
\ee
such that the matrix $\hat{P}^0_i(t)$ with elements 
\bea
P^0_{i; \{n\}\{m\}}(t) &=& \langle \Psi_0(t) \mid 
\, \mid i;\{m\}\rangle\langle i;\{n\}\mid\,\mid \Psi_0\rangle 
\nonumber\\
&=& \delta_{\{n\}\{m\}}\, \prod_\alpha 
\left(n^0_{i\alpha}(t)\right)^{n_\alpha}\,\left(1-n^0_{i\alpha}(t)\right)^{1-n_\alpha} \equiv
 \delta_{\{n\}\{m\}}\, P^0_{i;\{n\}}(t),
\label{tGA-P0-Fock}
\eea
is diagonal. We write a generic Gutzwiller projector as 
\be
\mathcal{P}_i(t) = \sum_{\Gamma\{n\}}\, 
\fract{\Phi_{i; \Gamma \{n\}}(t)}{\sqrt{P^0_{i;\{n\}}(t)}}\, 
\mid i;\Gamma\rangle\langle i;\{n\}\mid,\label{tGA-better-P}
\ee
with variational parameters $\Phi_{i; \Gamma\{n\}}(t)$ that define 
a matrix $\hat{\Phi}_i(t)$, and where $\mid i;\Gamma\rangle$ are basis states in the original representation 
in terms of the operators $c^\dagger_{ia}$. In fact, a nice feature of such a mixed original and natural basis 
representation of the Gutzwiller projectors is that one can carry out all calculations without specifying what 
the actual natural basis is;\cite{Nicola-HF,Rotationally,BG} it is just sufficient that this basis exists. 

In this representation, the constraints Eqs.~\eqn{1} and \eqn{2} can be simply rewritten
 as\cite{Nicola-HF} 
\bea
\Tr\Big(\hat{\Phi}_i(t)^\dagger\hat{\Phi}_i(t)^\dagger\Big) &=& 1,
\label{tGA-1}\\
\Tr\Big(\hat{\Phi}_i(t)^\dagger\hat{\Phi}_i(t)^\dagger
\hat{d}^\dagger_{i\alpha} \hat{d}^\dagga_{i\alpha}
\Big) &=& \langle \Psi_0(t)\mid d^\dagger_{i\alpha} d^\dagga_{i\alpha}
\mid\Psi_0(t)\rangle = n^0_{i\alpha}(t), \qquad \forall \alpha,
\label{tGA-2}\\
\Tr\Big(\hat{\Phi}_i(t)^\dagger\hat{\Phi}_i(t)^\dagger
\hat{d}^\dagger_{i\alpha} \hat{d}^\dagga_{i\beta}
\Big) &=& \langle \Psi_0(t)\mid d^\dagger_{i\alpha} d^\dagga_{i\beta}
\mid\Psi_0(t)\rangle = 0, \qquad \forall \alpha\not = \beta,
\label{tGA-3}
\eea
where, from now on, given any operator $\mathcal{O}_i$, we shall denote as $\hat{O}_i$ its representation in a basis of states . It turns out that only the constraint \eqn{tGA-2} requires some care to be implemented, while the other two can be implemented once for all at the beginning of the calculation.\footnote{In fact, we can parametrize 
\[
\hat{\Phi}_i(t) = \hat{U}_i(t) \, \sqrt{\hat{P}_i(t)},
\]
where $\hat{U}_i(t)$ is a unitary matrix with elements $U_{i\,\Gamma\{n\}}$, while 
$\hat{P}_i(t)$ a positive definite matrix with 
elements $P_{i\,\{n\}\{m\}}(t)$, which can be represented as the density matrix 
of a local normalized state 
\[
\mid \psi_i(t)\rangle = \sum_{\{n\}}\, c_{i\{n\}}(t) \mid i;\{n\}\rangle,
\]
with $\langle \psi_i(t)\mid \psi_i(t)\rangle = 1$, which automatically fulfills Eq.~\eqn{tGA-1}. 
In order to impose the constraint \eqn{tGA-2} it is then sufficient that, for $\alpha\not=\beta$  
\[
\langle \psi_i(t) \mid d^\dagger_{i\alpha}d^\dagga_{i\beta}\mid\psi_i(t)\rangle = 0.
\]
This can be done by regarding $\mid \psi_i(t)\rangle$ as the eigenstate of a local many-body Hamiltonian 
that does not contain any term of the form 
$c^\dagger_{i\alpha}\mid i;\{n\}\rangle\langle i;\{n\}\mid c^\dagga_{i\beta}$ for any $\mid \{n\}\rangle$ 
including the vacuum.
}

In this representation, Eq.~\eqn{O_i} becomes 
\be
\langle \Psi(t)\mid \mathcal{O}_i\mid \Psi(t)\rangle 
= \Tr\Big(\hat{\Phi}_i(t)^\dagger\,\hat{O}_i\,
\hat{\Phi}_i(t)\Big),\label{tGA-O_i}
\ee 
hence the average of any local operator can be expressed solely in terms 
of the matrices $\hat{\Phi}_i$ without any reference to the Slater 
determinant. In terms of $\hat{\Phi}_i$ one can show that 
\be
\langle\Psi_0(t)\mid \mathcal{P}_i(t)^\dagger 
\dot{\mathcal{P}}_i(t)^\dagga\mid\Psi_0(t)\rangle = 
\Tr\bigg(\hat{\Phi}_i(t)^\dagger 
\fract{\partial \hat{\Phi}_i(t)}{\partial t}\bigg).
\label{tGA-def:tPhi}
\ee

Also the effective Hamiltonian $\mathcal{H}_*(t)$ can be expressed simply in terms of the 
matrices $\hat{\Phi}_i(t)$. We define a matrix $\hat{R}_i(t)$ whose 
elements are \cite{Rotationally,Nicola-HF}
\be
R_{i\,a\alpha}(t) = \fract{1}{\sqrt{n^0_{i\alpha}(t)\left(1-n^0_{i\alpha}(t)\right)}}\; 
\Tr\Big(\hat{\Phi}_i(t)^\dagger 
\hat{c}^\dagga_{ia}\hat{\Phi}_i(t) \hat{d}^\dagger_{i\alpha}\Big),\label{tGA-def:R_ab}
\ee
which, by Eq.~\eqn{tGA-2}, can be regarded as functional of $\hat{\Phi}_i$ alone. In terms of 
those parameters, 
\be
\mathcal{H}_*\Big[\hat{\Phi}(t)\Big] = \sum_{i,j}\,\sum_{a,b=1}^N\,\sum_{\alpha,\beta=1}^N\, \Big(
 d^\dagger_{i\alpha} \, R_{i\,\alpha a}(t)^\dagger \,t^{ab}_{ij}\,R_{j\,b\beta}(t)\, d^\dagga_{j\beta} + H.c.
 \Big),
 \label{tGA-H*}
 \ee
and we must make sure that this non-interacting Hamiltonian does produces a local density matrix 
diagonal in the $d_{i\alpha}$ operators. 
In conclusion, having introduced the matrices $\hat{\Phi}_i$, we 
can rewrite the Lagrangian \eqn{tGA-L} as 
\bea
\mathcal{L}(t) 
&=& \sum_i\,i\Tr\bigg(\hat{\Phi}_i(t)^\dagger 
\fract{\partial \hat{\Phi}_i(t)}{\partial t}\bigg)
-\Tr\Big(\hat{\Phi}_i(t)^\dagger \hat{U}_i\,
\hat{\Phi}_i(t)\Big) \nonumber\\
&& + i\langle\Psi_0(t)\mid \dot{\Psi}_0(t)\rangle 
- \langle \Psi_0(t)\mid \mathcal{H}_*\left[\hat{\Phi}(t)\right]
\mid\Psi_0(t)\rangle.
\label{tGA-def:L}
\eea

We still need to impose the constraint Eq.~\eqn{tGA-2} in a convenient manner.  In fact, what we are going 
to show now is that we do not need to impose any constraint at time $t>0$ if that constraint is fulfilled 
at time $t=0$. Since the matrix $\hat{\Phi}_i$ is variational, we can always write 
\[
\hat{\Phi}_i \rightarrow \hat{\Phi}'_i\, \hat{V}_i^\dagger,
\]
with $\hat{\Phi}'_i$ and $\hat{V}_i^\dagger$ on the right hand side being independent variables. 
We assume that $\hat{V}_i$ is a unitary matrix that corresponds to a unitary operator $\mathcal{V}_i$ 
such that 
\bea
\mathcal{V}_i^\dagger d^\dagga_{i\alpha} \mathcal{V}_i &=& \sum_b V_{i\,\alpha\beta}\,
d^\dagga_{i\beta},
\label{tGA-V1}\\
\hat{V}_i^\dagger \hat{d}^\dagga_{i\alpha} \hat{V}_i &=& \sum_b V_{i\,\alpha\beta}\,
\hat{d}^\dagga_{i\beta}.
\label{tGA-V2}
\eea
It is straightforward to show that 
\be
\hat{R}_i\left[\hat{\Phi}_i\right] \rightarrow \hat{R}_i\left[\hat{\Phi}'_i\right] \,\hat{V}_i^\dagger,
\label{tGA-R-V}
\ee
so that 
\be
\mathcal{H}_*\left[\hat{\Phi}\right] \rightarrow \mathcal{V}\,\mathcal{H}_*\left[\hat{\Phi}'\right] \,
\mathcal{V}^\dagger,\label{tGA-H*-V}
\ee
where $\mathcal{V} = \prod_i \mathcal{V}_i$. Therefore the Lagrangian transforms into 
\bea
\mathcal{L}(t) 
&=& \sum_i\,i\Tr\bigg(\hat{\Phi}'_i(t)^\dagger 
\fract{\partial \hat{\Phi}'_i(t)}{\partial t}\bigg)
+ i \Tr\bigg(\hat{\Phi}'_i(t)^\dagger \hat{\Phi}'_i(t)\, 
\fract{\partial \hat{V}_i(t)^\dagger}{\partial t}\,\hat{V}_i(t)\bigg)
\nonumber\\
&& -\Tr\Big(\hat{\Phi}'_i(t)^\dagger \hat{U}_i\,
\hat{\Phi}'_i(t)\Big) \nonumber\\
&&  + i\langle\Psi_0(t)\mid \dot{\Psi}_0(t)\rangle 
- \langle \Psi_0(t)\mid \mathcal{V}\,\mathcal{H}_*\left[\hat{\Phi}'(t)\right]\,
\mathcal{V}^\dagger
\mid\Psi_0(t)\rangle.
\label{tGA-def:L-V}
\eea
Since also the Slater determinant is a variational parameter, we can redefine 
\[
\mid \Psi_0(t)\rangle \rightarrow \mathcal{V}\mid\Psi'_0(t)\rangle,
\]
where $\mid\Psi'_0(t)\rangle$ is still a Slater determinant, because of our definition of $\mathcal{V}$, and 
is independent of it. 
It follows that 
\bea
\mathcal{L}(t) 
&=& \sum_i\,i\Tr\bigg(\hat{\Phi}'_i(t)^\dagger 
\fract{\partial \hat{\Phi}'_i(t)}{\partial t}\bigg)
+ i \Tr\bigg(\hat{\Phi}'_i(t)^\dagger \hat{\Phi}'_i(t)\, 
\fract{\partial \hat{V}_i(t)^\dagger}{\partial t}\,\hat{V}_i(t)\bigg)
\nonumber\\
&& -\Tr\Big(\hat{\Phi}'_i(t)^\dagger \hat{U}_i\,
\hat{\Phi}'_i(t)\Big) \nonumber\\
&& + i\langle\Psi'_0(t)\mid \dot{\Psi}'_0(t)\rangle 
+ i\langle\Psi'_0(t)\mid \mathcal{V}(t)^\dagger\dot{\mathcal{V}}(t)\mid\Psi'_0(t)\rangle \nonumber\\
&& - \langle \Psi'_0(t)\mid \mathcal{H}_*\left[\hat{\Phi}'(t)\right]\,
\mid\Psi'_0(t)\rangle,
\label{tGA-def:L-V-bis}
\eea
where the only piece of the Lagrangian that depends explicitly on $\mathcal{V}$ is, being 
$\mathcal{V}$ unitary,   
\be
\delta\mathcal{L}\left[\hat{V}^\dagger,\hat{\dot{V}}\right] = - i \Tr\bigg(\hat{\Phi}'_i(t)^\dagger \hat{\Phi}'_i(t)\, 
\hat{V}_i(t)^\dagger \fract{\partial \hat{V}_i(t)}{\partial t}\,\bigg)
+ i\langle\Psi'_0(t)\mid \mathcal{V}(t)^\dagger\dot{\mathcal{V}}(t)\mid\Psi'_0(t)\rangle .
\label{tGA-L-V}
\ee

Now, let us assume that 
\be
\mathcal{V}_i(t) = \exp\bigg[ -i\sum_\alpha\,\phi_{i\alpha}(t)\,
d^\dagger_{i\alpha}d^\dagga_{i\alpha}\bigg].
\label{simple-def:V}
\ee
It follows that \eqn{tGA-L-V} becomes 
\be
\delta\mathcal{L}\left[\phi,\dot{\phi}\right] = \sum_\alpha \, \dot{\phi}_{i\alpha}(t)
\bigg[ - \Tr\bigg(\hat{\Phi}'_i(t)^\dagger \hat{\Phi}'_i(t)\, \hat{d}^\dagger_{i\alpha} 
\hat{d}^\dagga_{i\alpha}\bigg)
+ \langle\Psi'_0(t)\mid d^\dagger_{i\alpha}d^\dagga_{i\alpha}\mid\Psi'_0(t)\rangle \bigg].
\label{simple-L-V}
\ee
Since this is the only term that depends on $\phi_{ia}$, the Euler-Lagrange equation
\[
\fract{\partial \mathcal{L}}{\partial \phi_{ia}} 
- \fract{d}{dt}\, \fract{\partial \mathcal{L}}{\partial \dot{\phi_{ia}}} =0,
\]
implies that 
\bea
0 &=& \fract{d}{dt}
\bigg[ - \Tr\bigg(\hat{\Phi}'_i(t)^\dagger \hat{\Phi}'_i(t)\, \hat{d}^\dagger_{i\alpha} 
\hat{d}^\dagga_{i\alpha}\bigg)
+ \langle\Psi'_0(t)\mid d^\dagger_{i\alpha}d^\dagga_{i\alpha}\mid\Psi'_0(t)\rangle \bigg]\nonumber\\
&=& \fract{d}{dt}
\bigg[ - \Tr\bigg(\hat{\Phi}'_i(t)^\dagger \hat{\Phi}'_i(t)\, \hat{V}^\dagger_i(t)\,
\hat{d}^\dagger_{i\alpha} \hat{d}^\dagga_{i\alpha}V_i(t)\bigg)
+ \langle\Psi'_0(t)\mid \mathcal{V}(t)^\dagger d^\dagger_{i\alpha}d^\dagga_{i\alpha}
\mathcal{V}(t)\mid\Psi'_0(t)\rangle \bigg]\nonumber\\
&=& \fract{d}{dt}
\bigg[ - \Tr\bigg(\hat{\Phi}_i(t)^\dagger \hat{\Phi}_i(t)\, \hat{d}^\dagger_{i\alpha} 
\hat{d}^\dagga_{i\alpha}\bigg)
+ \langle\Psi_0(t)\mid d^\dagger_{i\alpha}d^\dagga_{i\alpha}\mid\Psi_0(t)\rangle \bigg].\label{simple-constraint}
\eea
In other words, provided Eq.~\eqn{tGA-2} is satisfied at $t=0$, and Eqs.~\eqn{tGA-1} and \eqn{tGA-3} 
are enforced by construction, then the constraint \eqn{tGA-2} is automatically satisfied by the saddle point solution at any time $t\geq 0$ . 

In conclusion, under the above assumptions, the only requirement is finding the saddle point of 
the action whose Lagrangian is given in Eq.~\eqn{tGA-def:L}. Specifically, the Slater determinant 
must satisfy the equation
\be
i\mid \dot{\Psi}_0(t)\rangle = \mathcal{H}_*\Big[\hat{\Phi}(t)\Big]\mid \Psi_0(t)\rangle,
\label{simple:Eq for Psi0}
\ee
which is just a Schr{\oe}dinger equation with a time-dependent Hamiltonian that depends parametrically 
on the matrices $\hat{\Phi}_i(t)$. These latter in turns must satisfy
\be
i \fract{\partial \hat{\Phi}_i(t)}{\partial t} = 
\hat{U}_i \hat{\Phi}_i(t) + 
\langle \Psi_0(t)\mid 
\fract{\partial \mathcal{H}_*\Big[\hat{\Phi}(t)\Big]}{\partial \hat{\Phi}_i(t)^\dagger}
\mid \Psi_0(t)\rangle \equiv \hat{H}_i\Big[\Psi_0(t),\hat{\Phi}(t)\Big]\,\hat{\Phi}_i(t)
,\label{simple:Eq for Phi}
\ee
which is a non-linear Schr{\oe}dinger equation whose Hamiltonian $\hat{H}_i$ 
depends not only on the Slater determinant $\mid \Psi_0(t)\rangle$ but also on the same 
$\hat{\Phi}_i(t)$ at site $i$ and on the $\hat{\Phi}_j(t)$'s at the neighboring sites.  We note that the time-evolution 
as set by the Eqs.~\eqn{simple:Eq for Psi0} and \eqn{simple:Eq for Phi} is unitary, hence conserves the 
energy if the Hamiltonian is not explicitly time dependent. In other words, one can readily show that 
\be
\fract{d E(t)}{d t} \equiv
 \fract{d}{d t} \langle \Psi(t)\mid \mathcal{H}\mid\Psi(t)\rangle = 
\fract{d}{d t} \langle \Psi_0(t)\mid \mathcal{H}_*\Big[\hat{\Phi}(t)\Big]
\mid\Psi_0(t)\rangle = 0,
\label{simple:unitary}
\ee
if $\mid \Psi_0(t)\rangle$ satisfies Eq.~\eqn{simple:Eq for Psi0}, while   
$\hat{\Phi}_i(t)$  and $\hat{\Phi}_i(t)^\dagger$  satisfy Eq.~\eqn{simple:Eq for Phi} and its hermitean 
conjugate, respectively. If $\mathcal{H}(t)$ is explicitly time-dependent then, under the same conditions as before,  
\be
\fract{d E(t)}{d t} \equiv
 \fract{d}{d t} \langle \Psi(t)\mid \mathcal{H}(t)\mid\Psi(t)\rangle = 
\langle \Psi_0(t)\mid \fract{\partial \mathcal{H}_*\Big[t,\hat{\Phi}(t)\Big]}{\partial t}
\mid\Psi_0(t)\rangle,
\label{simple:non-unitary}
\ee
where the time derivative in the r.h.s. only refers to the explicit time dependence.

The stationary limit of \eqn{simple:Eq for Psi0} and \eqn{simple:Eq for Phi}, i.e.
\bea
E\Big[\hat{\Phi}\Big] \mid\Psi_0\rangle &=& \mathcal{H}_*\Big[\hat{\Phi}\Big] \mid\Psi_0\rangle,
\label{simple-stationary-Psi0}\\
\Lambda\Big[\Psi_0\Big] \hat{\Phi}_i &=& 
\bigg( \hat{U}_i + \fract{\partial E\Big[\hat{\Phi}\Big]}{\partial \hat{\Phi}_i^\dagger}\bigg) 
\,\hat{\Phi}_i \equiv \hat{H}_i\Big[\Psi_0,\hat{\Phi}\Big]\,\hat{\Phi}_i,
\label{simple-stationary-Phi}
\eea
for the lowest eigenvalues $E$ and $\Lambda$ corresponds to solving the conventional equilibrium problem discussed in section \ref{Model}, as showed in Ref.~\cite{Nicola}. In particular, the Eq.~\eqn{simple-stationary-Phi} is a self-consistent eigenvalue equation similar to Hartree-Fock, in which the Hamiltonian depends parametrically on the same eigenstate that is looked for.

In conclusion, the Eqs.~\eqn{simple-stationary-Psi0} and \eqn{simple-stationary-Phi} for the stationary 
condition at equilibrium, and the Eqs.~\eqn{simple:Eq for Psi0} and \eqn{simple:Eq for Phi} for 
the out-of-equilibrium evolution, provide a very simple tool for studying the correlations effect in a strongly 
interacting electron model. The method is very flexible; it can deal with many orbitals and also 
with inhomogeneous situations where the Hamiltonian and/or the initial state are not 
translationally invariant, hence the matrices $\hat{\Phi}_i(t)$ become site dependent.  
We stress once more that the approach is {\sl variational} only in the limit 
of infinite lattice-coordination, otherwise it is just a mere approximation without any control parameter, 
exactly like DMFT when it is used in finite and not just in infinite dimensions. 

One aspect worth to be mentioned is that within the Gutzwiller approximation two different types of dynamical degrees of freedom seem to emerge. One is 
provided by the Slater determinant with its evolution \eqn{simple:Eq for Psi0}. It is commonly believed that 
this set just describes the quasiparticle degrees of freedom. In addition, the matrices $\hat{\Phi}_i$ introduce other local degrees of freedom with their own dynamics set by Eq.~\eqn{simple:Eq for Phi}. 
It is tempting to associate them with the incoherent excitations that 
coexist with the coherent quasiparticles in the presence of interaction, and which become the Hubbard bands 
near a Mott transition.\cite{DMFT} Within the Gutzwiller approximation, coherent and incoherent excitations are coupled to each other in a mean field like fashion, which provides 
a very intuitive picture although it misses important dissipative mechanisms. In what follows, we shall provide additional evidences that $\hat{\Phi}_i$ are indeed related to the Hubbard bands.

\subsection{A simple case study}
\label{A simple case study}

Before concluding this section, we think it is worth showing how the equation simplify in the frequent and relevant cases in which  the point symmetry of the Hamiltonian already determines  
the local orbitals in which representation the local single-particle density matrix is diagonal, i.e. 
\be
\langle \Psi(t)\mid c^\dagger_{ia} c^\dagga_{ib}\mid \Psi(t)\rangle = 
\delta_{ab}\,n_{ia}(t).\label{simple-1}
\ee
In this case, where natural and original basis coincide, hence also  
\be
\langle \Psi_0(t)\mid c^\dagger_{ia} c^\dagga_{ib}\mid \Psi_0(t)\rangle = \delta_{ab}
\,n^0_{ia}(t),\label{simple-n0}
\ee
 the expression \eqn{tGA-def:R_ab} further simplifies into 
\be
R^*_{i\,ab}(t) = \delta_{ab}\, \fract{1}{\sqrt{n^0_{ia}(t)\Big(1-n^0_{ia}(t)\Big)}}\;
\Tr\Big(\hat{\Phi}_i(t)^\dagger 
\hat{c}^\dagger_{ia}\hat{\Phi}_i(t) \hat{c}^\dagga_{ia}\Big)\equiv 
R^*_{ia}(t)\,\delta_{ab}.\label{simple-R_ab}
\ee
Because of the constraint Eq.~\eqn{tGA-2}, we can equivalently regard 
\be
n^0_{ia}(t) = \Tr\Big(\hat{\Phi}_i(t)^\dagger \hat{\Phi}_i(t)\,\hat{n}_{ia}\Big),
\label{simple-n0-Phi}
\ee
as functional of $\hat{\Phi}_i(t)$, rather than of the Slater determinant, hence it follows that 
\bea
\fract{\partial R^*_{ia}(t)}{\partial \hat{\Phi}_i(t)^\dagger} &=& 
\fract{1}{\sqrt{n^0_{ia}(t)\Big(1-n^0_{ia}(t)\Big)}}\; 
\hat{c}^\dagger_{ia}\hat{\Phi}_i(t) \hat{c}^\dagga_{ia}\nonumber\\ 
&& + R^*_{ia}(t)\,\fract{2n^0_{ia}-1}{2n^0_{ia}\Big(1-n^0_{ia}(t)\Big)}\;
\hat{\Phi}_i(t)\,\hat{n}_{ia},\label{simple-dR*}\\
\fract{\partial R_{ia}(t)}{\partial \hat{\Phi}_i(t)^\dagger} &=& 
\fract{1}{\sqrt{n^0_{ia}(t)\Big(1-n^0_{ia}(t)\Big)}}\; 
\hat{c}^\dagga_{ia}\hat{\Phi}_i(t) \hat{c}^\dagger_{ia} \nonumber\\
&& + R_{ia}(t)\,\fract{2n^0_{ia}-1}{2n^0_{ia}\Big(1-n^0_{ia}(t)\Big)}\;
\hat{\Phi}_i(t)\,\hat{n}_{ia}.\label{simple-dR}
\eea
If we consider the Hamiltonian 
\be
\mathcal{H} = \sum_{ij}\,\sum_{ab} t^{ab}_{ij}
\Big(c^\dagger_{ia}c^\dagga_{jb} + H.c.\Big) + \sum_i\, \mathcal{U}_i,\label{simple-Ham}
\ee
then 
\be
\mathcal{H}_*(t) = \sum_{ij}\,\sum_{ab} t^{ab}_{ij}\Big(R_{ia}(t)^*\,R_{jb}(t)\,
c^\dagger_{ia}c^\dagga_{jb} + H.c.\Big),\label{simple-H*}
\ee
so that, through \eqn{tGA-def:L}, the Slater determinant satisfies that Schr{\oe}dinger equation\footnote{
Once again, we must make sure that the effective Hamiltonian $\mathcal{H}_*(t)$, Eq.~\eqn{simple-H*}, 
is such that the local density matrix remains indeed diagonal in the operators $c^\dagger_{ia}$. } 
\be
i\mid \dot{\Psi}_0(t)\rangle = \mathcal{H}_*(t)\mid\Psi_0(t)\rangle.\label{simple-Psi0}
\ee

If we define 
\be
\Delta_{ia}(t) = \sum_{jb}\,t^{ab}_{ij}\,R_{jb}(t)\,
\langle \Psi_0(t)\mid c^\dagger_{ia}c^\dagga_{jb}\mid\Psi_0(t)\rangle,\label{simple-def:Delta}
\ee
then $\hat{\Phi}_i$ satisfies the matricial Schr{\oe}dinger equation
\bea
i \fract{\partial \hat{\Phi}_i(t)}{\partial t} &=& \hat{U}_i\,\hat{\Phi}_i(t) 
+ \sum_a\, \fract{\Delta_{ia}(t)}{\sqrt{n^0_{ia}(t)\Big(1-n^0_{ia}(t)\Big)}}\; 
\hat{c}^\dagger_{ia}\hat{\Phi}_i(t) \hat{c}^\dagga_{ia} \nonumber\\
&& ~~~~~~~~~~~ + \sum_a\, \fract{\Delta_{ia}(t)^*}{\sqrt{n^0_{ia}(t)\Big(1-n^0_{ia}(t)\Big)}}\; 
\hat{c}^\dagga_{ia}\hat{\Phi}_i(t) \hat{c}^\dagger_{ia} \nonumber\\
&&~~~~~~~~~~~~ + \sum_a\, 
\Big(R^*_{ia}(t)\,\Delta_{ia}(t)+c.c.\Big)\,\fract{2n^0_{ia}-1}{2n^0_{ia}\Big(1-n^0_{ia}(t)\Big)}\;
\hat{\Phi}_i(t)\,\hat{n}_{ia}.\label{simple-dPhi}
\eea
The equation for $\hat{\Phi}_i^\dagger$ can be obtained simply by the hermitean conjugate of 
\eqn{simple-dPhi}. We can readily demonstrate, through \eqn{simple-R_ab}, that 
\bea
i\fract{d}{dt}\Tr\Big(\hat{\Phi}_i(t)^\dagger \hat{\Phi}_i(t) \hat{n}_{ib}\Big) 
&=& \sum_a\, 
\fract{\Delta_{ia}(t)}{\sqrt{n^0_{ia}(t)\Big(1-n^0_{ia}(t)\Big)}}\; 
\Tr\Big(\hat{\Phi}_i(t)^\dagger \hat{c}^\dagger_{ia}\hat{\Phi}_i(t) 
\big[\hat{c}^\dagga_{ia},\hat{n}_{ib}\big]\Big)\nonumber\\
&& + \sum_a\, \fract{\Delta_{ia}(t)^*}{\sqrt{n^0_{ia}(t)\Big(1-n^0_{ia}(t)\Big)}}\; 
\Tr\Big(\hat{\Phi}_i(t)^\dagger \hat{c}^\dagga_{ia}\hat{\Phi}_i(t) 
\big[\hat{c}^\dagger_{ia},\hat{n}_{ib}\big]\Big)\nonumber\\
&=& \fract{\Delta_{ib}(t)}{\sqrt{n^0_{ib}(t)\Big(1-n^0_{ib}(t)\Big)}}\; 
\Tr\Big(\hat{\Phi}_i(t)^\dagger \hat{c}^\dagger_{ib}\hat{\Phi}_i(t) 
\hat{c}^\dagga_{ib}\Big)\nonumber\\
&& -  \fract{\Delta_{ib}(t)^*}{\sqrt{n^0_{ib}(t)\Big(1-n^0_{ib}(t)\Big)}}\; 
\Tr\Big(\hat{\Phi}_i(t)^\dagger \hat{c}^\dagga_{ib}\hat{\Phi}_i(t) 
\hat{c}^\dagger_{ib}\Big)\nonumber\\
&=& R_{ib}(t)^*\,\Delta_{ib}(t) - R_{ib}(t)\,\Delta_{ib}(t)^* \nonumber\\
&=& \sum_{ja} t^{ba}_{ij} \Big(R_{ib}(t)^* R_{ja}(t) 
\langle \Psi_0(t) \mid c^\dagger_{ib}c^\dagga_{ja}\mid\Psi_0(t)\rangle - c.c.\Big)\nonumber\\
&=& \langle \Psi_0(t)\mid \Big[n_{ib}\,,\mathcal{H}_*(t)\Big]\mid \Psi_0(t)\rangle\nonumber\\
&=& i\fract{d}{dt}\,\langle \Psi_0(t)\mid n_{ib}\mid\Psi_0(t)\rangle
,\label{simple-dn-Phi}
\eea
which explicitly proves that the constraint is indeed conserved by the above dynamical evolution.  

\section{Quantum quenches in the half-filled Hubbard model}
\label{Quantum quenches in the half-filled Hubbard model}

Armed with all previous results, we can start investigating the simplest possible out-of-equilibrium 
evolution in the single-band Hubbard model at 
half-filling. For sake of simplicity we shall ignore magnetism, hence assume spin $SU(2)$ invariant 
$\mid \Psi_0(t)\rangle$ and $\hat{\Phi}_i$. In this case, natural and original bases coincide, hence we 
can use the results of section \ref{A simple case study}. We choose as a local basis that of 
an empty site, $\mid 0\rangle$, doubly-occupied site,  $\mid 2\rangle$, and 
singly occupied site with spin up, $\mid \up \rangle$, or down, $\mid \down\rangle$. 
We take for $\hat{\Phi}_i$ with elements $\Phi_{i\,\Gamma\Gamma'}$ with 
$\Gamma,\Gamma'=0,2,\up,\down$ the $SU(2)$ and particle-hole invariant form 
\be
\hat{\Phi}_i = \fract{1}{\sqrt{2}}
\begin{pmatrix}
\Phi_{i\,00} & 0 & 0 & 0\\
0 & \Phi_{i\,22} & 0 & 0\\
0 & 0 & \Phi_{i\,\up\up} & 0 \\
0 & 0 & 0 & \Phi_{i\,\down\down}
\end{pmatrix},\label{quench-Phi}
\ee
with $\Phi_{i\,00}=\Phi_{i\,22}\equiv \Phi_{i0}$ and 
$\Phi_{i\,\up\up} = \Phi_{i\,\down\down}\equiv \Phi_{i1}$. All constraints Eqs.~\eqn{tGA-1}-\eqn{tGA-3}, with $n^0_{i\up}(t)=n^0_{i\down}(t)=1/2$ $\forall t$, are satisfied provided 
\be
\mid \Phi_{i\,0}\mid^2 + \mid \Phi_{i\,1}\mid^2 = 1.\label{quench-constraint}
\ee
With the above parametrization the Eq.~\eqn{simple-R_ab} becomes
\be
R_{i\up}(t)^* = R_{i\down}(t)^* \equiv R_i(t)^* = 
\Phi_{i0}(t)^*\Phi_{i1}(t) + \Phi_{i1}(t)^*\Phi_{i0}(t)\in \mathbb{R}\text{e}.\label{quench-R}
\ee

Given the original Hamiltonian 
\be
\mathcal{H} = \sum_{ij\,\sigma}\, t_{ij}\,
\Big(c^\dagger_{i\sigma}c^\dagga_{j\sigma}+ H.c.\Big) + \fract{U}{2}\sum_i\,\Big(n_i-1\Big)^2,
\label{quench-Ham}
\ee
then 
\be
\mathcal{H}_*(t) = \sum_{ij\,\sigma}\, t_{ij}\,
\Big(R_i(t)\, R_j(t) \,c^\dagger_{i\sigma}c^\dagga_{j\sigma}+ H.c.\Big) ,
\label{quench-H*}
\ee
and the Slater determinant is the solution of the Schr{\oe}dinger equation \eqn{simple-Psi0}.
In this case in which $a=\up,\down$ and spin symmetry is preserved, the parameter defined 
in Eq.~\eqn{simple-def:Delta}
\be
\Delta_{i\up}(t) = \Delta_{i\down}(t) = \fract{\Delta_i(t)}{2} = \frac{1}{2}
\sum_{j\sigma}\, t_{ij}\,R_j(t) \,\langle \Psi_0(t)\mid c^\dagger_{i\sigma}c^\dagga_{j\sigma}
\mid\Psi_0(t)\rangle \in \mathbb{R}\text{e},\label{quench-Delta}
\ee 
is real. Therefore the equation of motion \eqn{simple-dPhi} becomes
\bea
i \dot{\Phi}_{i0}(t) &=& \frac{U}{2}\,\Phi_{i0}(t) 
+ 2\,\Delta_i(t)\,\Phi_{i1}(t),\label{quench-eq-0}\\
i \dot{\Phi}_{i1}(t) &=& 2\,\Delta_i(t)\,\Phi_{i0}(t).\label{quench-eq-1}
\eea
We note that if we imagine the spin-1/2 wave-function
\be
\mid \Phi_i(t)\rangle = \Phi_{i1}(t)\mid \Uparrow\rangle + \Phi_{i0}(t) \mid\Downarrow\rangle 
\ee
solution of the Schr{\oe}dinger equation of the spin Hamiltonian 
\be
\mathcal{H}_{*\text{Ising}} = \sum_i\, \fract{U}{4}\Big(1-\sigma^z_i\Big) 
+ 2\Delta_i(t)\,\sigma^x_i,\label{quench-Ising-Ham}
\ee
that describes independent spins in a uniform magnetic field $-U/4$ along $z$ and a site and time dependent 
field $2\Delta_i(t)$ along $x$, we would get exactly the equations \eqn{quench-eq-0} and 
\eqn{quench-eq-1}, with 
\be
R_i(t) = \langle \Phi_i\mid \sigma^x \mid \Phi_i\rangle,\label{quench-R.vs.sigma}
\ee
implying that the field $2\Delta_i(t)$ is self-consistently determined by the same spins. This observation 
is not a  coincidence, as we shall discuss later. 

Before analyzing a simple case of out-of-equilibrium evolution, let us consider the stationary 
limit, which, as we discussed, defines the equilibrium conditions. In this case it is likely that the lowest energy state is homogeneous, namely invariant under translations, hence $R_i=R$, $\forall i$. The stationary solution of Eq.~\eqn{simple-Psi0} is just the ground state of the hopping Hamiltonian with renormalized 
hopping parameters $t_{*\,ij} = R^2 t_{ij}$ and energy per site $R^2 \epsilon_0<0$. Therefore 
$\Delta_i = \Delta = R\,\epsilon_0$, for all $i$, hence the Eqs.~\eqn{quench-eq-0} and 
\eqn{quench-eq-1} in the stationary limit become simply (we drop the site index as all sites are equivalent)
\bea
\Lambda \, \Phi_{0} &=& \frac{U}{2}\,\Phi_{0} 
+ 2\,\epsilon_0\,R\,\Phi_{1},\label{stationary-eq-0}\\
\Lambda\,\Phi_{1} &=& 2\,\epsilon_0\,R\,\Phi_{0}.\label{stationary-eq-1}
\eea
We write $\Phi_0 = \sin\theta/2$ and $\Phi_1=\cos\theta/2$, so that the wavefunction is normalized, 
hence $R=\sin\theta$. The eigenvalue problem is solved if 
\be
\cos\theta = \fract{U}{8|\epsilon_0|},\label{stationary-theta}
\ee
for $U\leq U_c= 8 |\epsilon_0|$, in which case 
\[
\Lambda = \frac{U}{4} + 2\epsilon_0 = \frac{U}{4} - 2|\epsilon_0|,
\]
otherwise, for $U>U_c$, the solution is $\theta=0$ with energy $\Lambda=0$. Indeed, $U_c$ 
can be identified  as the critical repulsion for the Mott transition within the Gutzwiller approximation, 
because, for $U>U_c$, $R=\sin\theta=0$, hence the hopping energy vanishes. We observe that 
the highest energy eigenvalue at self-consistency is 
\[
\Lambda' = \frac{U}{4} + 2|\epsilon_0|,
\]
for $U\leq U_c$, and $\Lambda'=U/2$ above, resembling much what we would expect for 
the location of the Hubbard bands. 

Let us come back to the out-of-equilibrium evolution, and suppose we start at $t=0$ from the ground state of the non-interacting Hamiltonian, which is just the ground state average of the hopping with 
energy per site $\epsilon_0<0$ introduced above, and total energy $E_0<0$. This corresponds to assuming that $\Phi_{i0}(0)=\Phi_{i1}(0)=1/\sqrt{2}$, hence $R_i(0)=1$, $\forall i$,  
and $\mid\Psi_0(0)\rangle$ being the uniform non-interacting Fermi sea. Since translational symmetry 
remains unbroken during the time evolution, $R_i(t)=R(t)$, $\forall i$ and $\forall t>0$. It follows that 
$\mathcal{H}_*(t)$ remains the same tight-binding Hamiltonian as at $t=0$, just renormalized by the 
overall factor $R(t)^2$. As a result,  the Slater determinant evolution is trivial, 
\be
\mid \Psi_0(t) \rangle = \text{e}^{-i E_0\int_0^t dt' R(t')^2}\mid\Psi_0(0)\rangle,
\ee
hence $\Delta_i(t) = R(t)\,\epsilon_0$. Therefore the equations \eqn{quench-eq-0} and 
\eqn{quench-eq-1} become for any site equal to 
\bea
i \dot{\Phi}_{0}(t) &=& \frac{U}{2}\,\Phi_{0}(t) 
+ 2\,\epsilon_0\,R(t)\,\Phi_{1}(t),\label{quench-eq-0-bis}\\
i \dot{\Phi}_{1}(t) &=& 2\,\epsilon_0\,R(t)\,\Phi_{0}(t),\label{quench-eq-1-bis}
\eea
with $R(t)=\Phi_1(t)^*\Phi_0(t) + c.c.$. If we set
\bea
\langle \Phi(t)\mid \sigma^x\mid \Phi(t)\rangle &=& \sin\theta(t)\,\cos\frac{\phi(t)}{2},\label{quench-x}\\
\langle \Phi(t)\mid \sigma^y\mid \Phi(t)\rangle &=& \sin\theta(t)\,\sin\frac{\phi(t)}{2},\label{quench-y}\\
\langle \Phi(t)\mid \sigma^z\mid \Phi(t)\rangle &=& \cos\theta(t),\label{quench-z}
\eea
then, through Eqs.~\eqn{quench-eq-0-bis} and \eqn{quench-eq-1-bis}, we find the following 
equation of motion for $\phi(t)$:
\be
\dot{\phi}(t) = \pm \sqrt{ U^2 - 16\,\epsilon_0^2 \,\sin^2 \phi(t)},\label{quench-eq-phi}
\ee
which is just the equation of a pendulum. In particular, if $U\leq 4|\epsilon_0|$, $\phi(t)$ 
oscillates between $\pm \phi_{\text{Max}}$, where
\[
\phi_{\text{Max}} = \sin^{-1} \fract{U}{4|\epsilon_0|}.
\]
On the contrary, when $U>4|\epsilon_0|$, $\phi(t)$ increases indefinitely. In other words, the quench 
dynamics displays a dynamical critical point at $U_* = 4|\epsilon_0|$.\cite{Marco-PRL} 
We observe that $U_*$ is just one half of the critical $U_c$ that we found previously at the Mott transition within the Gutzwiller approximation. 
Remarkably, an abrupt change of dynamical behavior near $U_c/2$ has been observed also 
in Ref.~\cite{Werner_prl09} within a time-dependent DMFT simulation of the same quantum quench as above. Given the very crude approximation in using a Gutzwiller wavefunction with respect to the 
exactness of DMFT in infinite coordination lattices, such an agreement is indeed quite remarkable.


\section{A multi-configurational Gutzwiller approach}

In section \ref{Time-dependent Gutzwiller approximation} we already noticed that the 
variational degrees of freedom introduced by the projectors $\mathcal{P}_i$ are promoted to the rank of 
true dynamical degrees of freedom in the time dependent extension of the Gutzwiller approximation. 
Moreover, in section \ref{Quantum quenches in the half-filled Hubbard model} we found that 
in the simple case of a single-band Hubbard model at half-filling, these new dynamical objects 
resemble spins in a self-consistent magnetic field, see Eq.~\eqn{quench-Ising-Ham}. 
In what follows we will put such an analogy on a more solid basis, although the demonstration 
applies rigorously only to few simple cases. The outcome will be a theory that looks 
similar to the so-called slave-spin representation recently 
introduced\cite{DeMedici-1,DeMedici-2,Z2-1,Z2-2} as an alternative approach to 
slave-boson theory.

\subsection{$SU(N)$ Hubbard model at half-filling}

We note that, at given $\mid \Psi_0\rangle$, the Gutzwiller wave-function $\mid \Psi\rangle$ 
in Eq.~\eqn{Psi} actually defines a whole set of wave-functions, 
each identified by the projectors $\mathcal{P}_i$ that act on each site $i$. 
Let us assume there exist  a whole set of projectors 
$\mathcal{P}_{i\, m}$ that satisfy
\bea
\langle \Psi_0\mid \mathcal{P}^\dagger_{i\,m}
\mathcal{P}^\dagga_{i\,n}\mid\Psi_0\rangle &=& \delta_{nm},
\label{1-ext}\\
\langle \Psi_0\mid \mathcal{P}^\dagger_{i\,m}
\mathcal{P}^\dagga_{i\,n} c^\dagger_{ia\sigma}c^\dagga_{ib\sigma'} \mid\Psi_0\rangle &=& 
\delta_{mn}\, \langle \Psi_0\mid c^\dagger_{ia\sigma}c^\dagga_{ib\sigma}\mid \Psi_0\rangle, \;
\forall a,b \text{~and~} \forall \sigma,\sigma',
\label{2-ext}
\eea
where we distinguish between orbital indices, $a,b=1.\dots,N$, and spin indices, $\sigma$ and $\sigma'$.
It is straightforward realizing that these conditions allow to evaluate, along the same lines previously outlined, also matrix elements between 
different wave-functions. In this way, one can get the matrix representation of the Hamiltonian on such a subspace of wave-functions, whose diagonalization provides not only a better estimate of the ground state energy but also gives access to excited states. 

The Hamiltonian we shall consider is given by \eqn{Ham} with diagonal nearest neighbor hopping 
$-\delta_{ab}\,t/\sqrt{z}$ and 
\be
\mathcal{U}_i = \frac{U}{2}\,\Big(n_i-N\Big)^2,
\label{Hubbard-U}
\ee
where $n_i=\sum_{a\sigma} c^\dagger_{ia\sigma}c^\dagga_{ia\sigma}$, and  
the density corresponds to $N$ electrons per site, i.e. half-filling. The model therefore is invariant 
not only under spin $SU(2)$ but also orbital $SU(N)$, in fact it is invariant under the large $U(2N)$ 
symmetry group. We shall therefore assume that the wave functions $\mid\Psi\rangle$ and 
$\mid \Psi_0\rangle$ are invariant under such a large symmetry. We define $\mathcal{Q}_{in}$ 
the projection operator at site $i$ onto states with $n$ electrons. If we choose as local basis the Fock states $\mid i;\left\{n\right\}\rangle$ identified by the occupation numbers $n_{ia\sigma}=0,1$ 
in each orbital and spin, i.e. 
\[
\mid i;\left\{n\right\}\rangle = \prod_{a=1}^N\,\prod_\sigma\,\Big(c^\dagger_{ia\sigma}\Big)^{n_{ia\sigma}}\mid 0\rangle,
\]
then   
\[
\mathcal{Q}_{in} = \sum_{\{n_{ia\sigma}\}}\, \delta\bigg(n-\sum_{a\sigma}n_{ia\sigma}\bigg)\,
\mid i;\left\{n\right\}\rangle\langle i;\left\{n\right\}\mid.
\]
From the invariance properties of the Slater determinant $\mid\Psi_0\rangle$ it follows that 
\[
\langle \Psi_0\mid c^\dagger_{ia\alpha}c^\dagga_{ib\beta}\mid\Psi_0\rangle = \frac{1}{2}\;\delta_{ab}\delta_{\alpha\beta},
\]
as well as that 
\be
\langle \Psi_0\mid \mathcal{Q}_{in}\mid \Psi_0 \rangle = \fract{1}{4^N}\,\begin{pmatrix} 2N \\ n\end{pmatrix}
\equiv P^{(0)}_{n}=P^{(0)}_{2N-n},
\ee
where $P^{(0)}_{n}$ is the distribution probability of the local occupation number on the uncorrelated wavefunction.
The most general Gutzwiller projector satisfying \eqn{1} and \eqn{2} 
can be written as 
\be
\mathcal{P}_i = \sum_{n=0}^{2N}\,\fract{\Phi_{i\,n-N}}
{\sqrt{P^{(0)}_n}}\;
\mathcal{Q}_{in},\label{def:Pi}
\ee 
where 
\[
\sum_{s=-N}^{N}\,\mid\Phi_{i\,s}\mid^2 =1,
\]
and $\mid\Phi_{i\, s}\mid = \mid\Phi_{i\, -s}\mid$. In fact, we can regard $\Phi_{i\, s}$ as the wavefunction components of fictitious spins of magnitude $S=N$, one at each at site $i$,
\[
\mid \Phi_i\rangle = \sum_{s=-S}^S\, \Phi_{i\,s}\mid s\rangle_i, 
\]
which we shall intentionally denote as {\sl slave spins} as they are closely related to the slave-spin representations of Hubbard-like models.\cite{DeMedici-1,DeMedici-2,Z2-1,Z2-2}

The renormalization factor defined by Eq.~\eqn{def:R_ab} is in this case diagonal, $R_{i\, ab}= R_i\,\delta_{ab}$, and simply given by 
\bea
R_i^* &=& \sum_{s=-S}^{S-1}\, \Phi_{i\,s+1}^*\Phi_{i\,s}^\dagga\, 
\frac{1}{S}\;\sqrt{S(S+1) -s(s+1)}\nonumber\\
&=& \frac{1}{S}\;\langle \Phi_i\mid S^+\mid\Phi_i\rangle.\label{R-Phi}
\eea
More generally, the matrix element of the fermionic creation 
operator $c^\dagger_{ia\sigma}$ between two wave-functions, 
$\mid \Psi\rangle$ and $\mid \Psi'\rangle$, with local projectors $\mathcal{P}_i$ and $\mathcal{P}_i'$ at site $i$, hence 
slave spin wave functions $\mid \Phi_i\rangle$ and 
$\mid\Phi_i'\rangle$, respectively, has the very transparent expression 
\be
\mathcal{P}_i^\dagger c^\dagger_{ia\sigma} \mathcal{P}'_i 
\rightarrow \fract{\langle \Phi_i\mid S^+\mid \Phi_i'\rangle}{S}\;
c^\dagger_{ia\sigma}.\label{def:R-diff-Phi}
\ee   
Seemingly, the matrix element of the local repulsion reads
\be
\frac{U}{2}\,\langle \Psi\mid \Big(n_i-N\Big)^2\mid \Psi'\rangle 
= \frac{U}{2}\;\langle \Phi_i\mid \left(S^z\right)^2 \mid \Phi_i'\rangle,
\label{def:U-diff-Phi}
\ee
where $S^z$ is the $z$-component of the slave spin operator $\mathbf{S}$.
In conclusion, we find that 
\bea
\langle \Psi\mid\mathcal{H}\mid\Psi'\rangle &=& 
-\fract{t}{S^2\sqrt{z}}\,\sum_{<i,j> \sigma a}\,\bigg(
\langle \Phi_i\mid S^+\mid\Phi_i'\rangle 
\langle \Phi_j\mid S^-\mid\Phi_j'\rangle 
\langle \Psi_0\mid c^\dagger_{ia\sigma}c^\dagga_{ja\sigma}\mid\Psi_0\rangle + H.c.\bigg)\nonumber\\
&& + \frac{U}{2}\sum_i\, \langle \Phi_i\mid \left(S^z\right)^2 \mid \Phi_i'\rangle,\label{E-diff-Phi}
\eea
indeed a very suggestive result. Notice, however, that the slave spin wave-functions are not completely free, 
because they must correspond to Gutzwiller projectors satisfying \eqn{1-ext} and \eqn{2-ext}. 

\subsection{Slave-spin basis}
Therefore, to make Eq.~\eqn{E-diff-Phi} suitable for calculations, 
we still need to identify a proper set of Gutzwiller projectors satisfying Eqs.~\eqn{1-ext}, \eqn{2-ext}. 
 A possible choice is 
\bea
\mathcal{P}_{i\,0} &=& \sqrt{\fract{1}{P^{(0)}_N}}\;
\mathcal{Q}_{iN},\label{def:P0}\\
\mathcal{P}_{i\,m>0} &=& \sqrt{\fract{1}{2P^{(0)}_{N+m}}}\;
\Big(\mathcal{Q}_{iN+m}+\mathcal{Q}_{iN-m}\Big),\label{def:Pm}
\eea
with $m\leq N$. In principle we could have also chosen the combination 
\eqn{def:Pm} with the minus sign instead of the plus, but not both as they are not orthogonal in the sense of Eq.~\eqn{2-ext}. In other words, 
not the whole slave-spin Hilbert space is allowed, but only a subspace 
$\mid(m)\rangle$, with $m=0,\dots,S$:
\bea
\mid (0) \rangle &\equiv& \mid 0 \rangle,\label{0}\\
\mid (m>0) \rangle &\equiv& \fract{1}{\sqrt{2}}\,\big(\mid m\rangle + \mid -m \rangle\big),\label{m}
\eea
which we shall denote as the {\sl physical subspace}, still to keep contact with the jargon of slave-boson 
theories.

We note that the action of the raising operator $S^+$ projected 
onto the physical subspace, i.e. 
\bea
S^+\mid (0)\rangle &\simeq& \sqrt{\fract{S(S+1)}{2}}\;\mid (1)\rangle,\label{A}\\
S^+\mid (1)\rangle &\simeq&\sqrt{\fract{S(S+1)}{2}}\;\mid (0)\rangle
+ \sqrt{\fract{S(S+1)-2}{4}}\;\mid (2)\rangle,\label{B}\\
S^+\mid (m>1)\rangle &\simeq& 
\sqrt{\fract{S(S+1)-m(m+1)}{4}}\;\mid (m+1)\rangle \nonumber \\
&& + \sqrt{\fract{S(S+1)-m(m-1)}{4}}\;\mid (m-1)\rangle,\label{C}
\eea
is just the same as the action of $S^x$ without any restriction. Indeed
\ba
S^x \mid(0)\rangle &=& S^x\mid 0\rangle = \fract{\sqrt{S(S+1)}}{2}\,\Big(
\mid 1\rangle + \mid -1\rangle\Big) =   \sqrt{\fract{S(S+1)}{2}}\;\mid (1)\rangle,\\
S^x \mid(1)\rangle &=& \frac{1}{2\sqrt{2}}\bigg( 
\sqrt{S(S+1)-2}\;\mid 2\rangle + \sqrt{S(S+1)}\;\mid 0 \rangle \\
&& ~~~~~~~~ + \sqrt{S(S+1)}\;\mid 0 \rangle + \sqrt{S(S+1)-2}\;\mid -2\rangle \bigg)\\
&=& \sqrt{\fract{S(S+1)}{2}}\;\mid (0)\rangle + \sqrt{\fract{S(S+1)-2}{4}}\;\mid (2)\rangle,\\
\ea
and also \eqn{C} follows trivially. Actually, the {\sl physical} subspace is invariant under the action of $S^x$, therefore, 
using the latter instead of $S^+$, we are allowed to release the constraint and work in the full Hilbert 
space of the {\sl slave} spins, since we expect the ground state to contain $\mid 0\rangle$, hence to occur within the {\sl physical} subspace of Eqs.~\eqn{0} and \eqn{m}. 

In conclusion, we can rewrite \eqn{E-diff-Phi} as 
\bea
\langle \Psi\mid\mathcal{H}\mid\Psi'\rangle &=& 
-\fract{t}{S^2\sqrt{z}}\,\sum_{<i,j> \sigma a}\,
\langle \Phi_i\mid S^x\mid\Phi_i'\rangle 
\langle \Phi_j\mid S^x\mid\Phi_j'\rangle\,
\langle \Psi_0\mid c^\dagger_{ia\sigma}c^\dagga_{ja\sigma}
+ H.c.\mid\Psi_0\rangle \nonumber\\
&& + \frac{U}{2}\sum_i\, \langle \Phi_i\mid \left(S^z\right)^2 \mid \Phi_i'\rangle,\label{E-diff-Phi-final}
\eea  
without any condition to be imposed on the slave spin wave-functions. 
We finally note that Eq.~\eqn{E-diff-Phi-final} is just a matrix element of  the Hamiltonian 
\be
\mathcal{H}_* = -\fract{t}{S^2\sqrt{z}}\, 
\sum_{<i,j>\sigma a}\, S^x_i S^x_j\Big(c^\dagger_{ia\sigma}
c^\dagga_{ja\sigma} + H.c.\Big) + 
\frac{U}{2}\,\sum_i\, \left(S^z_i\right)^2,\label{slave-spin-H}
\ee
which describes electrons coupled to slave spins of magnitude 
$S=N$. In this representation the slave spins are not subject to any 
constraint. 

We note that the Hamiltonian \eqn{slave-spin-H} resembles much the slave-rotor representation for 
the multi-orbital Hubbard model of Ref.~\cite{Slave-Rotors}, with however a major difference. 
In fact the Hamiltonian \eqn{slave-spin-H} possesses only a discrete $Z_2$ gauge symmetry, unlike 
the slave-rotor Hamiltonian that has a larger $U(1)$ gauge symmetry. This difference has some important 
consequences that we discuss below.  

\subsection{The Mott transition}
The great advantage of the representation \eqn{slave-spin-H} is to make the Mott transition accessible 
already within the mean field approximation. The simplest mean-field approach amounts to assume 
a factorized variational wave-function $\mid \Psi\rangle = \mid \text{electrons}\rangle \times 
\mid \text{slave-spins}\rangle$. The minimum energy is obtained by choosing $\mid \Psi_0\rangle$ the 
Fermi sea of a simple tight-biding Hamiltonian. If we define 
\[
- J \equiv 
-\fract{t}{V\sqrt{z}}\,\sum_{<i,j> \sigma a}\,
\langle \Psi_0\mid c^\dagger_{ia\sigma}c^\dagga_{ja\sigma}
+ H.c.\mid\Psi_0\rangle,
\]
the hopping energy per site of the state $\mid\Psi_0\rangle$, then the slave-spin wavefunction 
must be the ground state of the Hamiltonian 
\be
\mathcal{H}_\text{Ising} = -\frac{J}{S^2}\;\frac{2}{z}\sum_{<i,j>}\, 
S^x_i S^x_j + \frac{U}{2}\,\sum_i\, \left(S^z_i\right)^2.
\label{H-Ising}
\ee
This spin Hamiltonian has a discrete $Z_2$ symmetry $S^x_i\to -S^x_i$, $\forall i$, which is 
spontaneously broken at small $U/J$, i.e. $\langle S^x_i\rangle$ is non-zero and corresponds to 
the order parameter, and restored only above a quantum critical point.  This Ising-like transition 
corresponds to the Mott transition in the original interacting model. In fact, the physical electron 
$c^\dagger_{i\sigma}$ translates in the model \eqn{slave-spin-H} into the composite operator 
$S_i^x\,c^\dagger_{i\sigma}$ hence, within mean-field, the long distance density matrix
\[
\lim_{|i-j|\to \infty} \langle c^\dagger_{i\sigma} c^\dagga_{j\sigma}\rangle 
\Rightarrow \lim_{|i-j|\to \infty} \langle S^x_i\,c^\dagger_{i\sigma} S^x_j\, c^\dagga_{j\sigma}\rangle 
= \lim_{|i-j|\to \infty} \langle  S^x_i S^x_j\rangle \; 
\langle c^\dagger_{i\sigma} c^\dagga_{j\sigma}\rangle.
\]
The average over the electron wave function, which is the ground state of the hopping, is long ranged. 
Therefore the long distance behavior of the physical electron density matrix depends critically on 
the slave-spin correlation function. In the symmetry broken phase, 
\[
\lim_{|i-j|\to \infty} \langle  S^x_i S^x_j \rangle 
\to \langle S^x\rangle^2 \not = 0,
\]
hence the physical electron density matrix is long ranged, as we expect in a metallic phase. On the 
contrary, when the symmetry is restored, then $\langle S^x_i S^x_j\rangle$ vanishes exponentially 
for $|i-j|\to\infty$, transferring such an exponential decay to the physical electron density matrix, which 
therefore does not describe anymore a metal phase but rather a Mott insulating one. 
It is important to notice that, in the actual slave-spin model \eqn{slave-spin-H}, a finite 
order parameter $\langle S^x_i\rangle$ corresponds to a phase with broken $Z_2$ gauge symmetry, 
which is possible in spite of the Elitzur's theorem\cite{Elitzur} because we are working in the 
limit of infinite lattice coordination.\cite{Maslanka,Baruselli} We also observe that in the symmetry broken phase 
there are not Goldstone modes because the symmetry is discrete, unlike what predicted by the slave-rotor 
mean field theory,\cite{Slave-Rotors} where these gapless modes are expected and associated with the 
zero-sound. 

The location of the Ising critical point of the slave-spin Hamiltonian \eqn{slave-spin-H} can be  determined approximately by assuming that it occurs for large enough $U$'s so that  it is safe to keep only states with $S^z=0,\pm 1$. 
We denote 
\ba
\mid \up \rangle &=& \mid 0\rangle,\\
\mid \down\rangle &=& \fract{1}{\sqrt{2}}\,\Big(\mid +1\rangle + \mid -1\rangle\Big),
\ea
as the two states of an Ising variable, and introduce Pauli matrices in this subspace. We find that 
the operator $S^x$ in this subspace acts like $\sigma^x\,\sqrt{S(S+1)/2}$, while $\left(S^z\right)^2$ 
like $\left(1-\sigma^z\right)/2$, so that \eqn{H-Ising} can be rewritten as
\be
 \mathcal{H}_\text{Ising} \simeq -J\,\frac{S(S+1)}{2S^2}\;\frac{2}{z}\sum_{<i,j>}\, 
\sigma^x_i \sigma^x_j + \frac{U}{4}\,\sum_i\, \left(1-\sigma^z_i\right),\label{H-Ising-2}
\ee
namely like a simple Ising model in a transverse field. We note that for the case $S=1$, 
the Hamiltonian \eqn{H-Ising-2} coincides with the slave-spin 
representation of the single-band Hubbard model.\cite{DeMedici-1,Z2-1,Z2-2,Marco-PRB} 

The model \eqn{H-Ising-2} has indeed a quantum phase transition 
that separates a ferromagnetic phase, $\langle\sigma^x_i\rangle\not = 0$, for small $U/J$, 
from a paramagnetic one, $\langle\sigma^x_i\rangle= 0$, for large $U/J$. This transition is actually the Mott transition in the slave-spin language and, within mean-field, it would occur at a critical 
\be
U_c \simeq 8J\, \frac{S(S+1)}{2S^2}.\label{Uc}
\ee 
We note that in the single-band case, $S=1$, $U_c$ coincides with the value obtained previously. 

Apart from making the Mott transition accessible by mean-field, the effective slave-spin model also 
uncover new dynamical excitations that it is natural to associate with the Hubbard bands. Indeed, the models \eqn{H-Ising} and its simplified version \eqn{H-Ising-2} display a spin-wave branch that becomes soft only 
at the transition. For very large $U$, 
the excitation energy becomes $\sim U/2$, just the location of the Hubbard bands. Needless to say, 
the mean field dynamics of these spins corresponds to the dynamics of the matrices $\hat{\Phi}_i$ 
that we introduced previously. 

\subsection{Away from half-filling}

We can repeat all the above calculations even away from half-filling. In this case, the slave-spin 
wave-functions $\mid \Phi_\alpha\rangle$ in the physical subspace must satisfy the conditions
\bea
\langle \Phi_\alpha\mid  \Phi_\beta\rangle = \delta_{\alpha\beta},\label{1-tris}
\langle \Phi_\alpha\mid S^z \mid \Phi_\beta\rangle = \delta\,\delta_{\alpha\beta},\label{2-tris}
\eea
where $\delta=n-N$ is the doping away from half-filling. The expression of the variational energy is 
modified into 
\bea
\langle \Psi\mid\mathcal{H}\mid\Psi'\rangle &=& 
-\fract{t}{\left(S^2-\delta^2\right)\sqrt{z}}\,\sum_{<i,j> \sigma a}\,\bigg(
\langle \Phi_i\mid S^+\mid\Phi_i'\rangle 
\langle \Phi_j\mid S^-\mid\Phi_j'\rangle 
\langle \Psi_0\mid c^\dagger_{ia\sigma}c^\dagga_{ja\sigma}\mid\Psi_0\rangle + H.c.\bigg)\nonumber\\
&& + \frac{U}{2}\sum_i\, \langle \Phi_i\mid \left(S^z\right)^2 \mid \Phi_i'\rangle.\label{E-diff-Phi-delta}
\eea
As before we need to identify the physical subspace for the slave-spins. 

The simplest case is when the average occupancy $n$ is integer, hence 
$\delta$ is integer, too, which requires more than a single band, i.e. $N>1$. 
Let us further assume $U$ large, so that we can just focus on the two {\sl physical} states 
\ba
\mid \up\rangle &\equiv& \mid \delta \rangle,\\
\mid \down\rangle &\equiv& \frac{1}{\sqrt{2}}\,\Big( \mid \delta+1\rangle + \mid \delta-1\rangle\Big),
\ea
which corresponds to the assumption that a kind of particle-hole symmetry is recovered 
close to the Mott transition.The raising operator projected onto this subspace has the action
\ba
S^+\mid \up\rangle &\simeq& \sqrt{\fract{S(S+1)-\delta(\delta+1)}{2}}\;\mid\down\rangle
\equiv \big(\alpha+\beta\big)\mid\down\rangle,\\
S^+\mid \down\rangle &\simeq & \sqrt{\fract{S(S+1)-\delta(\delta-1)}{2}}\;\mid\up\rangle
\equiv \big(\alpha-\beta\big)\mid\up\rangle,
\ea
hence  $S^+\simeq \alpha \sigma^x -i\beta\sigma^y$, with $\alpha>\mid\beta\mid$. 
It follows that the Ising variables are 
described by the effective Hamiltonian  
\be
 \mathcal{H}_\text{Ising} \simeq -J\,\frac{1}{S^2-\delta^2}\;\frac{2}{z}\sum_{<i,j>}\, 
\Big(\alpha^2\,\sigma^x_i \sigma^x_j +\beta^2\,\sigma^y_i \sigma^y_j\Big)
+ \frac{U}{4}\,\sum_i\, \left(1-\sigma^z_i\right),\label{H-Ising-3}
\ee 
with $J$ being the average hopping per site of the Fermi sea with average occupation $n$. 
This model still has a phase transition between a ferromagnetic phase with $\langle \sigma^x\rangle\not = 0$ 
and a paramagnetic one. Within mean field, the critical interaction strength is now
\be
U_c  \simeq 8J \fract{\alpha^2}{S^2-\delta^2},\label{Uc-doping}
\ee
and is shifted to lower values 
of the interaction as $\delta$ increases. 
Once again, the spin-wave spectrum of the Ising model \eqn{H-Ising-3} can be 
interpreted as the spectrum of the Hubbard bands. 

If the filling is not an integer or the enlarged $SU(2N)$ symmetry is lowered, the above construction does not work anymore because we cannot define in general more than a single Gutzwiller projector satisfying both \eqn{1-ext} and \eqn{2-ext}.  In other words, 
while for integer fillings and $SU(2N)$ symmetry we can associate the dynamical variables $\hat{\Phi}_i$ with auxiliary spin operators, which allows for instance to improve the Gutzwiller approximation by including systematically quantum fluctuations, away from such a high-symmetry points we are unable to make such a simple identification, hence we must limit our analysis to the mean field dynamics of 
$\hat{\Phi}_i$.   

\section{Conclusions}

In this paper we have shown in detail how one can access by simple means the out-of-equilibrium time evolution of  a Gutzwiller-type variational wave function. The approach is rigorously variational in the limit of large coordination numbers, otherwise can be regarded as the dynamical counterpart of the widely adopted Gutzwiller approximation. The method is really simple to implement and very flexible. It is apt to cope with 
weak non-equilibrium compatible with linear response, but also with strong out-of-equilibrium conditions like sudden quantum quenches. It can describe single- and multi-band systems, as well as homogeneous and 
inhomogeneous models. 

The key feature that distinguishes the present method from the conventional time-dependent Hartree-Fock is 
the emergence of two distinct types of excitations that control the time-evolution of the wave function. 
One corresponds to the particle-hole excitations of the guiding Slater determinant, just like 
in the time-dependent Hartree-Fock, and is supposed to describe coherent quasiparticles. 
In addition, new local dynamical degrees of freedom emerge, which can be associated with the 
Hubbard bands and that are promoted to the rank of genuine excitations with their own dynamics. Within the Gutzwiller approach the Hubbard bands and the quasiparticles are mutually coupled in a mean-field 
like fashion, i.e. each of them generates a time-dependent field that acts on the other. In spite of such an 
approximation, the dynamical behavior that follows is quite richer than in Hartree-Fock. We have shown just 
an example of such a richness, namely the dynamical transition that occurs in the single-band Hubbard model at half-filling after a sudden increase of the repulsion.\cite{Marco-PRL} 

Finally, we have shown that it is possible to extend the variational approach to a multi-configurational 
wave function that comprises a linear combination of orthogonal 
Gutzwiller-type of wave functions. Such a multi-configurational variational method can be worked out 
analytically only in specific cases, specifically for integer fillings. Nevertheless it is quite instructive 
since it demonstrates that the above discussed time-dependent Gutzwiller approach is nothing but 
the mean-field approximation applied to the actual Hamiltonian dynamics within that subspace of 
orthogonal Gutzwiller wave functions. Remarkably, the Hamiltonian projected in that subspace resembles 
the slave-spin representations of correlated electron models,\cite{DeMedici-1,DeMedici-2,Z2-1,Z2-2}
thus providing a very intuitive picture of these theories.   
\begin{acknowledgement}
These proceedings are based on the work that I have done in collaboration with Marco Schir\`o, whom I thank warmly. I am also grateful to Nicola Lanat\`a for useful discussions. I also acknowledge support by the EU under the project GOFAST. 
\end{acknowledgement}


\end{document}